\documentclass[showpacs,prb,altafilletter,runinadress,reprint,twocolumn]{revtex4-1}

\usepackage{graphicx,psfrag,epsf,epsfig}
\usepackage{dcolumn}
\usepackage{bm,bbm}
\usepackage{pstricks,pst-plot}     
\usepackage{latexsym}
\usepackage{mathrsfs,amsmath,amssymb,textcomp}
\usepackage{hyperref}

\def\d{{\mathrm d}}
\newcommand{\pf}[1]{\frac{\partial}{\partial #1}}

\begin{document}


\title{Nonequilibrium Green's function theory of coherent excitonic effects in the
photocurrent response of semiconductor nanostructures}

\author{U. Aeberhard}
\email{u.aeberhard@fz-juelich.de}

\affiliation{IEK-5: Photovoltaik, Forschungszentrum J\"ulich, D-52425 J\"ulich,
Germany}

\date{\today}

\begin{abstract}
Excitonic contributions to absorption and photocurrent generation in semiconductor nanostructures
are described theoretically and simulated numerically using steady-state non-equilibrium Green's
function theory. In a first approach, the coherent interband polarization including Coulomb
corrections is determined from a Bethe-Salpeter-type equation for the equal time interband 
single-particle charge carrier Green's function. The effects of excitonic absorption on photocurrent
generation are considered on the same level of approximation via the derivation of the corresponding
corrections to the electron-photon self-energy.
\end{abstract} 

\pacs{71.35.-y, 72.20.Jv, 72.40.+w, 73.21.Fg, 73.40.Kp, 78.67.De }
\maketitle

\section{\label{sec:level1}Introduction} 
Recently, the investigation of semiconductor nanostructures for
photovoltaic applications has been of ever growing interest. Potential
candidates among these low dimensional absorbers are ordered configurations of quantum wells (QW)
\cite{ned:99_2,green:00} or quantum dots (QD) \cite{conibeer:06_tsf,marti:06}, which are widely used
in other optoelectronic devices such as lasers or light-emitting devices. However, due to the unique
operating regime of solar cells, optical and transport properties are equally important, and should not be considered in
the isolated nanostructure component, but for an open system connected to the environment
via contacts. Since the main attraction of the systems is the presence of
quantum confinement effects that can be exploited to enhance the photovoltaic
performance, a comprehensive description should be on the level of a
quantum transport theory. Such a theory was recently developed on the example
of QW solar cells \cite{ae:prb_08,ae:nrl_11,ae:jcel_review} and applied to QD solar cells
\cite{ae:oqe_12}.

Excitonic effects play only a minor role in conventional inorganic bulk
semiconductor solar cells, since in most cases, the exciton binding energies are small as compared
to the thermal broadening at room temperature, and exciton dissociation is very fast
as nothing hinders the spatial separation of the carriers. However, this is
not the case in quantum confined systems, where the strong localization of the
electron and hole wave functions leads to a large overlap and thus substantially larger
exciton binding energies. As a consequence, the excitonic features in the
optoelectronic properties persist up to room temperature and have therefore
considerable impact on the photovoltaic properties of devices based on such
systems. In the past, excitonic effects in semiconductor nanostructures have
been discussed for steady state linear absorption or in the regime of high-intensity pulse
excitation.
For the latter, sophisticated quantum-kinetic theories were developed
\cite{haug:84,haug:88,haug:92,haug:96,haug:04,henneberger:86,henneberger:88,henneberger:88_2,henneberger:88_3,henneberger:96,jahnke:95}.
For the description of quantum photovoltaic devices, the picture of coherent
excitonic absorption needs to be combined with a steady state quantum transport
formalism. A suitable theoretical framework is provided by the
non-equilibrium Green's function formalism. However, the shifting of the focus
from transient interband kinetics to steady-state transport does not allow for a
straight-forward inclusion of excitonic processes: while in the former
case coherent excitonic polarization can be included via the Fock
term of Coulomb interaction to lowest order, there is no equal-time
approximation in the latter situation. This paper thus aims at the inclusion of excitonic effects into a
general theory of quantum opto-electronics including the transport aspect, which should allow for
the study of photovoltaic systems where these effects dominate the photocurrent response
close to the absorption edge.

The paper is organized as follows. In the section after this introduction, the coupling of
charge carriers to the coherent radiation fields is described based on the NEGF theory
for a two-band  model of a direct gap semiconductor, followed by a derivation of
the coherent interband polarization and the effective interband self-energy due
to Coulomb-enhanced electron-photon interaction. In a further section,
these results are used in the simulation of the photocurrent respose of bulk and
thin film devices, where the latter case is represented by a single quantum well III-V
semiconductor $p$-$i$-$n$ diode.

\section{NEGF model of a contacted excitonic absorber}
\subsection{Hamiltonian, Green's functions and self energies}
As a suitable model system, we choose a simple two band model of a direct gap
semiconductor nanostructure selectively connected to ohmic contacts
\cite{ae:prb_08} and coupled to a coherent external photon field, which at this stage is treated
classically\footnote{The description of spontaneous emission would require the additional coupling
to an incoherent internal photon field.}. Since we are interested in the photocurrent response of
the system, only the electronic part of the system is considered via the Hamiltonian 
\begin{align}
\hat{H}=\hat{H}_{0}+\hat{H}_{e\gamma}+\hat{H}_{ep}+\hat{H}_{ee}+\hat{H}_{C}.
\end{align}
$\hat{H}_{0}$ is the Hamiltonian of the non-interacting isolated
mesoscopic absorber, $\hat{H}_{e\gamma}$ is the light-matter coupling, $\hat{H}_{ep}$ encodes
the electron-phonon and $\hat{H}_{ee}$ the carrier-carrier interaction. The last term $\hat{H}_{C}$
describes the (selective) coupling to contacts required for carrier extraction in order to enable photocurrent flow.
The charge carriers in the two bands $b=c,v$ are described by the field operators
$\hat{\Psi}_{b}(\mathbf{r},t)$, which provides the Hamiltonian representation
\begin{align}
\mathcal{H}(t)=\sum_{a,b=c,v}\int d^{3}r
\hat{\Psi}_{a}^{\dagger}(\mathbf{r},t)\hat{H}(\mathbf{r},t)\hat{\Psi}_{b}(\mathbf{r},t).
\end{align}
The renormalizing effects of the interaction and contact Hamiltonian terms on the isolated system
are expressed within non-equilibrium many-body perturbation theory \cite{kadanoff:62,keldysh:65} in
terms of corresponding self-energies entering the generalized Kadanoff-Baym equations for the charge
carrier non-equilibrium Green's functions
($\underline{1}\equiv\{\mathbf{r}_{1},t_{1}\in\mathcal{C}\},\mathcal{C}$: Keldysh contour in the
complex plane, $\int \d\underbar{1}\equiv \int d^{3}r_{1}\int_{\mathcal{C}}dt_{1}$)
\begin{align}
\mathbf{G}_{0}^{-1}(\underline{1},\underline{1})\mathbf{G}(\underline{1},\underline{1'})
=&\boldsymbol{\delta}(\underline{1},\underline{1'})\nonumber\\
&+\int\d\underbar{2}\,
\boldsymbol{\Sigma}(\underline{1},\underline{2})\mathbf{G}(\underline{2},\underline{1'}),
\label{eq:diffdys1}\\ 
[\mathbf{G}_{0}^{\dagger}]^{-1}(\underline{1'},\underline{1'})
\mathbf{G}(\underline{1},\underline{1'})=
&\boldsymbol{\delta}(\underline{1},\underline{1'})\nonumber\\&+ \int\d \underbar{2}
\,\mathbf{G}(\underline{1},\underline{2})
\boldsymbol{\Sigma}(\underline{2},\underline{1'}),\label{eq:diffdys2}
\end{align}
where 
\begin{equation}
[G_{0}^{-1}(\underline{1},\underline{1'})]_{ab}=\left(i\hbar\pf{t_{1}}-[H_{0}(\mathbf{r}_{1})]_{a}\right)
\delta(\underline{1}, \underline{1'})\delta_{ab}
\end{equation}
and the contour-ordered Green's functions are defined via
\begin{equation}
G_{ab}(\underline{1},\underline{1'})\equiv-\frac{i}{\hbar} \langle
\hat{\Psi}_{a}(\underline{1})\hat{\Psi}_{b}^{\dagger}(\underline{1}')\rangle_{\mathcal{C}}
\end{equation}
for band indices $a,b$. The self-energy term in the above equations for the Green's functions may be
divided into the contributions from the interactions and the contact term,
\begin{align}
 \boldsymbol{\Sigma}(\underline{1},\underline{2})=\boldsymbol{\Sigma}^{I}(\underline{1},\underline{2})
 +\boldsymbol{\Sigma}^{C}(\underline{1},\underline{2}),
\end{align}
where the interaction term contains the effects of electron-photon, electron-phonon and
electron-electron coupling,
\begin{align}
\boldsymbol{\Sigma}^{I}(\underline{1},\underline{2})=\boldsymbol{\Sigma}^{e\gamma}(\underline{1},\underline{2})
+\boldsymbol{\Sigma}^{ep}(\underline{1},\underline{2})+\boldsymbol{\Sigma}^{ee}(\underline{1},\underline{2}).
\end{align}
Following the standard real-time decomposition rules
\cite{langreth:76} applied to \eqref{eq:diffdys1} and a special band decoupling procedure described
in App.~\ref{app:decoupling}, the equations for the retarded, advanced, lesser and greater
components of the non-equilibrium Green's functions for charge carriers can be written in the
standard intraband form used in transport calculations, ($1\equiv
\{\mathbf{r}_{1},t_{1}\in\mathbbm{R}\},~\int \d 1 \equiv \int
d^{3}r_{1}\int_{-\infty}^{\infty}dt_{1}$)
\begin{align}
&\int \d 2\,\left[G^{-1}_{0,aa}(1,2)-\tilde{\Sigma}_{aa}^{R}(1,2)\right]G_{aa}^{R}(2,1')=\delta(1,1'),\nonumber\\
&G_{aa}^{<}(1,1')=\int\d 2\int\d 3\,
G_{aa}^{R}(1,2)\tilde{\Sigma}_{aa}^{<}(2,3)G_{aa}^{A}(3,1'),\nonumber\\
&\qquad\qquad\qquad a\in \{c,v\}\label{eq:dyson_intra}
\end{align} 
where 
\begin{align}
\tilde{\Sigma}_{aa}^{i}(1,1')=&\Sigma_{aa}^{i}(1,1')+\Sigma_{aa}^{\delta i}(1,1'),\nonumber\\&
i=R,A,\lessgtr, \quad a\in \{c,v\},
\end{align}
with the effective band-coupling self-energy from the singular contributions given by
\begin{align}
\Sigma_{aa}^{\delta i}(1,1')=&\int \d 2\int \d 3\,\Sigma_{ab}^{\delta}(1,2)
\tilde{G}_{bb}^{i}(2,3)\Sigma_{ba}^{\delta}(3,1'),\nonumber\\&
i=R,A,\lessgtr,\quad a\neq b, \quad a,b\in \{c,v\}.
\end{align}
In the situation under investigation, the singular interband self-energy itself is of the form
\begin{align}
\Sigma_{ab}^{\delta}(1,2)=&\left[\Sigma_{ab}^{\delta,e\gamma}(\mathbf{r}_{1},t_{1})
\delta(\mathbf{r}_{1}-\mathbf{r}_{2})
+\Sigma_{ab}^{\delta,ee}(\mathbf{r}_{1},\mathbf{r}_{2},t_{1})\right]\nonumber \\
&\times\delta(t_{1}-t_{2}),
\end{align}
where $\Sigma_{ab}^{\delta,e\gamma}$ encodes the coupling of electrons to a coherent photon field
and $\Sigma_{ab}^{\delta,ee}$ is the (non-retarded) Fock term of the
(screened) Hartree-Fock approximation to carrier-carrier interaction leading to
Coulomb enhancement of the optical interband transitions due to the electron-hole coupling.

\subsection{Singular self-energy and coherent interband polarization}
The self-energy due to the light-matter interaction can be written in terms of
the vector potential of the classical electromagnetic field,
\begin{align} 
\Sigma_{ab}^{\delta,e\gamma}(\mathbf{r},t)&=-\frac{
e}{m_{0}}\langle \hat{\mathbf{A}}(\mathbf{r},t)\rangle\cdot
\hat{\mathbf{p}}(\mathbf{r}),\label{eq:singular_se_phot}
\end{align}
with $\langle \hat{\mathbf{A}}(\mathbf{r},t)\rangle=\mathbf{A}_{coh}(\mathbf{r},t)$.
The Coulomb term is
\begin{align}
\Sigma_{ab}^{\delta,ee}(\mathbf{r}_{1},\mathbf{r}_{2},t)=i\hbar
V(\mathbf{r}_{1}-\mathbf{r}_{2})G_{ab}^{<}(\mathbf{r}_{1},\mathbf{r}_{2},t,t^{+}),\label{eq:cb_sing_se}
\end{align}
where $V$ is the (screened) Coulomb potential, and depends thus on the coherent
interband polarization through the interband Green's function, which the decoupling provides in the form
\begin{align}
G_{vc}^{<}(1,1')=&\int d
2\int
d3\,\Big[\tilde{G}_{vv}^{R}(1,2)\Sigma_{vc}^{\delta}(2,3)
G_{cc}^{<}(3,1')\nonumber\\&+\tilde{G}_{vv}^{<}(1,2)
\Sigma_{vc}^{\delta}(2,3) G_{cc}^{A}(3,1')\Big]\\ 
\equiv&-\frac{i}{\hbar}\int d 2\int
d3\,\Sigma_{vc}^{\delta}(2,3)\mathcal{T}_{vc}(1,2,3,1'),\label{eq:decoupling_2}
\end{align}
where we have defined
\begin{align}
\mathcal{T}_{vc}(1,2,3,1')\equiv&
i\hbar\Big[\tilde{G}_{vv}^{R}(1,2) 
G_{cc}^{<}(3,1')\nonumber\\&+\tilde{G}_{vv}^{<}(1,2)
G_{cc}^{A}(3,1')\Big].
\end{align} 
Under steady state conditions, Fourier-transform to the energy domain yields
\begin{align} 
\tilde{\Sigma}_{aa}^{i}(\mathbf{r}_{1},\mathbf{r}_{2};E)
=&\int d^3 r_{2}\int d^3 r_{3}\int \frac{dE'}{2\pi\hbar}
\Sigma_{ab}^{\delta}(\mathbf{r}_{1},\mathbf{r}_{2};E')\nonumber\\
&\times
\tilde{G}_{bb}^{i}(\mathbf{r}_{2},\mathbf{r}_{3};E-E')
\Sigma_{ba}^{\delta}(\mathbf{r}_{3},\mathbf{r}_{1}';E'),\nonumber\\&
\quad i=R,A,\lessgtr, \label{eq:eff_se}
\end{align}
with the singular self-energies given by corresponding Fourier transforms of Eqs. \eqref{eq:singular_se_phot} and
\eqref{eq:cb_sing_se}. Inserting the explicit expressions for the latter leads to a
Bethe-Salpeter type equation (BSE) for the coherent polarization, which for steady state
reads
\begin{widetext}
\begin{align}
G_{vc}^{<}(\mathbf{r}_{1},\mathbf{r}_{1'};E)=&-\frac{i}{\hbar}\int
d^{3}r_{2}\int d^{3}r_{3}\left[-\frac{
e}{m_{0}}\hat{\mathbf{A}}_{coh}(\mathbf{r}_{2},E)\cdot
\hat{\mathbf{p}}(\mathbf{r}_{2})\delta(\mathbf{r}_{2}-\mathbf{r}_{3})+i\hbar
V(\mathbf{r}_{2}-\mathbf{r}_{3})G_{vc}^{<}(\mathbf{r}_{2},\mathbf{r}_{3},E)\right]\nonumber\\
&\times\mathcal{T}_{vc}(\mathbf{r}_{1},\mathbf{r}_{2},\mathbf{r}_{3},\mathbf{r}_{1'};E)\\
&\equiv G_{vc,(0)}^{<}(\mathbf{r}_{1},\mathbf{r}_{1'};E)+\int d^{3}r_{2}\int
d^{3}r_{3}V(\mathbf{r}_{2}-\mathbf{r}_{3})\mathcal{T}_{vc}(\mathbf{r}_{1},\mathbf{r}_{2},
\mathbf{r}_{3},\mathbf{r}_{1'};E) G_{vc}^{<}(\mathbf{r}_{2},\mathbf{r}_{3},E),\label{eq:BSE_realspace}
\end{align}
\end{widetext}
with
\begin{align}
G_{vc,(0)}^{<}(\mathbf{r}_{1},\mathbf{r}_{1'};E)=&\frac{ie}{m_{0}\hbar}\int
d^{3}r_{2}\hat{\mathbf{A}}_{coh}(\mathbf{r}_{2},E)\cdot
\hat{\mathbf{p}}(\mathbf{r}_{2})\nonumber\\&\times\mathcal{T}_{vc}(\mathbf{r}_{1},
\mathbf{r}_{2},\mathbf{r}_{2},\mathbf{r}_{1'};E)\\
\equiv&-\frac{i}{\hbar}\int d^{3}r_{2}\int
d^{3}r_{3}\Sigma_{ab}^{\delta,e\gamma}(\mathbf{r}_{2};E)\nonumber\\
&\times\delta(\mathbf{r}_{2}-\mathbf{r}_{3})\mathcal{T}_{vc}(\mathbf{r}_{1},
\mathbf{r}_{2},\mathbf{r}_{3},\mathbf{r}_{1'};E)\label{eq:polfun_coh}
\end{align}
the coherent interband polarization of non-interacting electron-hole pairs, where 
\begin{align}
&\mathcal{T}_{vc}(\mathbf{r}_{1},\mathbf{r}_{2},\mathbf{r}_{3},\mathbf{r}_{4};E) =i\hbar\int
\frac{dE'}{2\pi\hbar}\Big[\tilde{G}_{vv}^{R}(\mathbf{r}_{1},\mathbf{r}_{2};E'-E)\nonumber\\
&\times G_{cc}^{<}(\mathbf{r}_{3},\mathbf{r}_{4};E')+\tilde{G}_{vv}^{<}(\mathbf{r}_{1},\mathbf{r}_{2};E'-E)	
G_{cc}^{A}(\mathbf{r}_{3},\mathbf{r}_{4};E')\Big].
\end{align}
Here, it is interesting to note that
$\mathcal{T}_{vc}(\mathbf{r}_{1},\mathbf{r}_{2},\mathbf{r}_{2},\mathbf{r}_{1};E)\equiv
\mathcal{P}_{cv}^{R}(\mathbf{r}_{1},\mathbf{r}_{2};E)$ is the retarded
component of the random-phase approximation of the incoherent polarization function used to
describe the interband coupling that is not singular in time \cite{ae:prb_11}.
The microscopic, non-local interband susceptibility $\chi_{vc}$ is introduced via
\begin{align}
G_{vc}^{<}(\mathbf{r}_{1},\mathbf{r}_{1'};E)=&\frac{i}{\hbar}\int
d^{3}r_{2}\int 
d^{3}r_{3}~\hat{\mathbf{d}}_{vc}(\mathbf{r}_{2}-\mathbf{r}_{3})\cdot
\boldsymbol{\mathcal{E}}(\mathbf{r}_{2},E)\nonumber\\
&\times\mathcal{\chi}_{vc}(\mathbf{r}_{1},
\mathbf{r}_{2},\mathbf{r}_{3},\mathbf{r}_{1'};E),
\end{align}
where $\hat{\mathbf{d}}=-e\hat{\mathbf{r}}$ is the dipole operator. The linear
macroscopic interband susceptibility is obtained from the macroscopic interband polarization given
by \cite{schaefer:02}
\begin{align}
\mathbf{P}(\mathbf{r}_{1},E)=&-i\hbar \int d^{3}r_{1'}\Big[
\hat{\mathbf{d}}_{cv}(\mathbf{r}_{1}-\mathbf{r}_{1'}) G_{vc}^{<}(\mathbf{r}_{1},\mathbf{r}_{1'};E)\nonumber\\
&+\hat{\mathbf{d}}_{vc}(\mathbf{r}_{1}-\mathbf{r}_{1'}) G_{cv}^{<}(\mathbf{r}_{1},\mathbf{r}_{1'};E)\Big]\\
\equiv& \int d^{3}r_{1'}\overleftrightarrow{\boldsymbol{\chi}}(\mathbf{r}_{1},\mathbf{r}_{1'};E)
\boldsymbol{\mathcal{E}}(\mathbf{r}_{1'},E),\label{eq:macroscopic_interbandpol}
\end{align}
i.e.,
\begin{align}
&\overleftrightarrow{\boldsymbol{\chi}}(\mathbf{r}_{1},\mathbf{r}_{1'};E)=\int d^{3}r_{2}\int
d^{3}r_{3}~ \Big[\hat{\mathbf{d}}_{cv}(\mathbf{r}_{1}-\mathbf{r}_{2})
\hat{\mathbf{d}}_{vc}(\mathbf{r}_{1'}-\mathbf{r}_{3})\nonumber\\&\times\mathcal{\chi}_{vc}(\mathbf{r}_{1},
\mathbf{r}_{1'},\mathbf{r}_{3},\mathbf{r}_{2};E)+\hat{\mathbf{d}}_{vc}(\mathbf{r}_{1}-\mathbf{r}_{2})
\hat{\mathbf{d}}_{cv}(\mathbf{r}_{1'}-\mathbf{r}_{3})\nonumber\\&\times
\mathcal{\chi}_{cv}(\mathbf{r}_{1},\mathbf{r}_{1'},\mathbf{r}_{3},\mathbf{r}_{2};E)\Big]
\end{align}
The susceptibility can be used to compute the \emph{local} absorption coefficient,
\begin{align}
&\alpha_{i}(\mathbf{r};E_{\gamma})=\frac{E_{\gamma}}{\hbar
\varepsilon_{0}c_{0}\sqrt{\varepsilon_{b}}}\Im \chi_{ii}(\mathbf{r},\mathbf{r};E_{\gamma}),~ i\in\{x,y,z\},\\
&\equiv
\frac{\hbar}{\varepsilon_{0}c_{0}\sqrt{\varepsilon_{b}}E_{\gamma}}\left(\frac{e}{m_{0}}\right)^2\lim_{\mathbf{r'}\rightarrow\mathbf{r}}\hat{p}_{i}(\mathbf{r})
\hat{p}_{i}(\mathbf{r'})G_{vc}^{<}(\mathbf{r},\mathbf{r}';E_{\gamma}),
\end{align}
where $c_{0}$ is the speed of light in vacuum and $\varepsilon_{b}$ is the background dielectric
constant. The corresponding \emph{average} (bulk) absorption coefficient
may then be defined via $\bar{\alpha}_{i}(E_{\gamma})\equiv \mathcal{V}^{-1}\int
d^{3}r\alpha_{i}(\mathbf{r};E_{\gamma})$\cite{haug:04}, with $\mathcal{V}$ the absorbing volume.

A BSE-type equation similar to \eqref{eq:BSE_realspace} can also be derived for the singular
self-energy using \eqref{eq:decoupling_2} in Eq. \eqref{eq:cb_sing_se},
\begin{align}
\Sigma_{ab}^{\delta}(\mathbf{r}_{1},\mathbf{r}_{1}',E)=&\Sigma_{ab,(0)}^{\delta}(\mathbf{r}_{1},\mathbf{r}_{1}',E)
+V(\mathbf{r}_{1}-\mathbf{r}_{1}')\nonumber\\&\times\int d^{3}r_{2}\int
d^{3}r_{3}\mathcal{T}_{ab}(\mathbf{r}_{1},
\mathbf{r}_{2},\mathbf{r}_{3},\mathbf{r}_{1'};E)\nonumber\\&\times\Sigma_{ab}^{\delta}(\mathbf{r}_{2},\mathbf{r}_{3},E),\quad
a\neq b\in \{c,v\},
\end{align}
where $\Sigma_{ab,(0)}^{\delta}=\Sigma_{ab}^{e\gamma,\delta}$. With the carrier Green's function modified by
the effective band coupling self-energy via Eq. \eqref{eq:dyson_intra}, the spectral response
$\mathcal{S}(\hbar\omega_{\gamma})\equiv{J}(\hbar\omega_{\gamma})/[e\phi_{\gamma}(\hbar\omega_{\gamma})]$, 	where
$\phi_{\gamma}$ denotes the spectral photon flux, is determined from the steady state current induced under monochromatic illumination in the interacting region, for which the standard
 Meir-Wingreen expression is used \cite{meir:92},
\begin{align}
&J(\hbar\omega_{\gamma})=\sum_{\alpha}S_{\alpha}^{-1}\int\frac{dE}{2\pi}\int d^{3}r\int
d^{3}r'\Big[\hbar^{-1} \Gamma^{\alpha}(\mathbf{r},\mathbf{r'};E)\nonumber\\&\times \{n_{F}(E-\mu_{\alpha})
A(\mathbf{r}',\mathbf{r};E)+i G^{<}(\mathbf{r}',\mathbf{r};E)\}\Big],\label{eq:mw}
\end{align}
where $S_{\alpha}$ is the surface area, $\Gamma^{\alpha}$ the broadening function and 
$\mu_{\alpha}$ the chemical potential of contact $\alpha$, $n_{F}$ is the Fermi function and $A\equiv
i(G^{R}-G^{A})$ the spectral function of the fully interacting and
contacted absorber, which here may be either bulk or a thin film of semiconductor material.

\section{Applications}

In this section, the self-consistent
band coupling self-energy approach derived above is first validated for the case of a contacted bulk
absorber and then implemented in the existing NEGF model for thin-film and quantum well solar cells
 \cite{ae:prb_08,ae:nrl_11}.

\subsection{Bulk}
For a periodic bulk material, the BSE for the interband polarization function can be rewritten in
Fourier space as follows:
\begin{align}
G^{<}_{vc}(\mathbf{k};E)=&~G^{<}_{vc,(0)}(\mathbf{k};E)\nonumber\\&+\mathcal{T}(\mathbf{k};E)\sum_{\mathbf{q}}
V(\mathbf{k}-\mathbf{q})G^{<}_{vc}(\mathbf{q};E)\\
=&
~G^{<}_{vc,(0)}(\mathbf{k};E)-\frac{i}{\hbar}\mathcal{T}(\mathbf{k};E)\Sigma^{\delta,Cb}_{vc}(\mathbf{k},E)\label{eq:BSE_bulk}\\
=&-\frac{i}{\hbar}\mathcal{T}(\mathbf{k};E)\Sigma^{\delta}_{vc}(\mathbf{k},E).
\end{align}
If the photon momentum is neglected as compared to the electron quasi-momentum, the non-interacting
polarization function reads
\begin{align}
G_{vc,(0)}^{<}(\mathbf{k};E)\approx&\frac{i
e}{m_{0}\hbar}\sum_{\mathbf{q}}\hat{\mathbf{A}}_{coh}(\mathbf{q},E)\cdot
\mathbf{p}_{vc}\mathcal{T}(\mathbf{k};E) \label{eq:polfun}
\end{align}
with
\begin{align}
\mathcal{T}(\mathbf{k};E) =&i\hbar\int
\frac{dE'}{2\pi\hbar}\Big[\tilde{G}_{vv}^{R}(\mathbf{k};E'-E)G_{cc}^{<}(\mathbf{k};E')\nonumber\\
&+\tilde{G}_{vv}^{<}(\mathbf{k};E'-E) G_{cc}^{A}(\mathbf{k};E')\Big].
\end{align}
To lowest order, inserting the quasi-equilibrium approximation for the bulk Green's
functions, 
\begin{align}
G_{aa}^{<}(\mathbf{k};E)=&2\pi if(E)\delta(E-\varepsilon_{a}(\mathbf{k})),\\
G_{aa}^{R/A}(\mathbf{k};E)=&\left[E-\varepsilon_{a}(\mathbf{k})\pm
i\eta\right]^{-1},\quad \eta\rightarrow 0^{+},
\end{align}
the following expression is obtained
\begin{align}
\mathcal{T}(\mathbf{k};E)
=\frac{f_v(\mathbf{k})-f_c(\mathbf{k})}{\varepsilon_{c}(\mathbf{k})-\varepsilon_{v}(\mathbf{k})-E+i\eta},
\end{align}
which, used in \eqref{eq:polfun}, leads to the standard form of the macroscopic
polarization function\cite{haug:04}. 

\begin{figure}[!b]
\includegraphics[width=8.5cm]{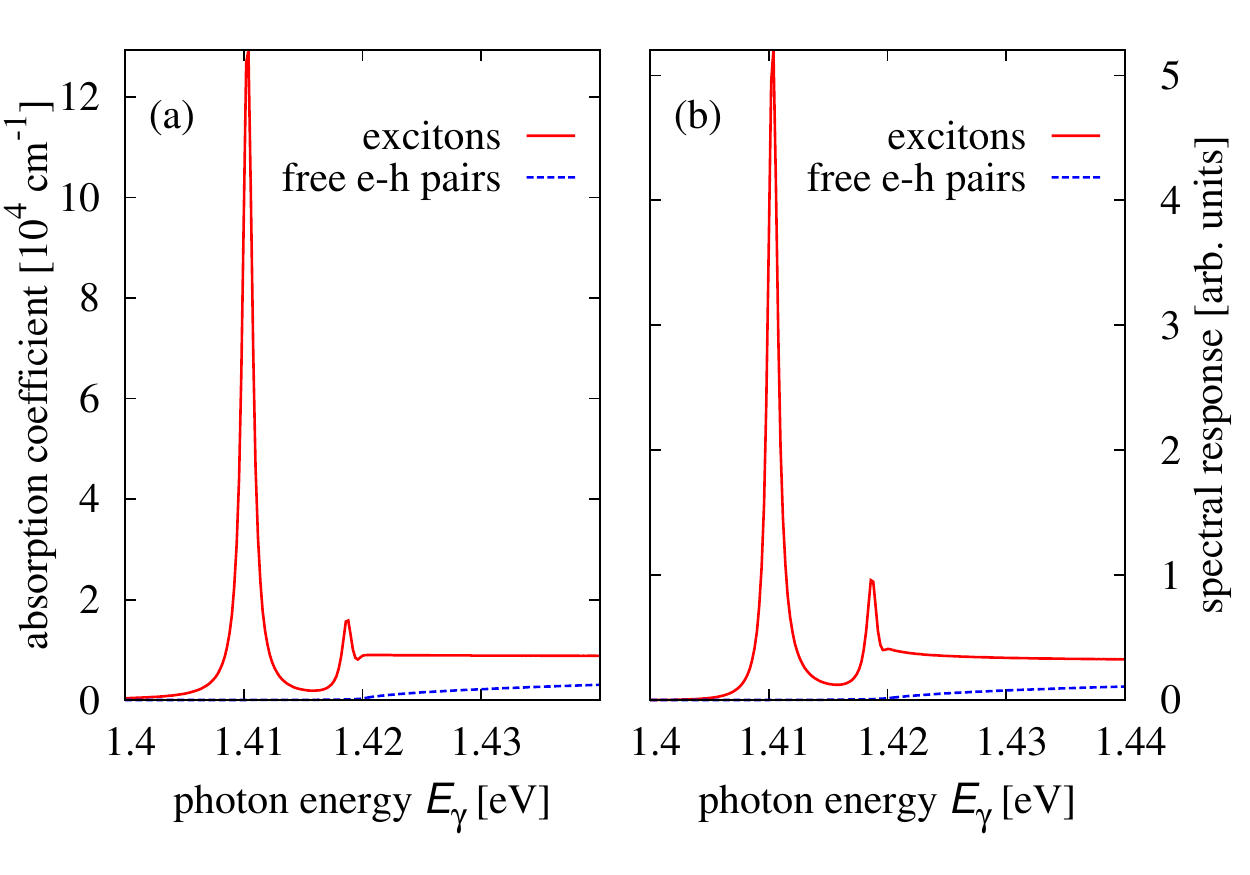}
\caption{(a) Linear absorption coefficient of a bulk direct gap semiconductor absorber described by
two parabolic bands, as computed via the coherent interband polarization function corrected for
electron-hole interactions. For comparison, the absorption of non-interacting electron-hole pairs
is shown as well. (b) Spectral response for the same system as computed via the effective
electron-photon self energy corrected for electron-hole interactions.
\label{fig:excitons_bulk}}
\end{figure}

Assuming complete isotropy, one may neglect the angular dependence and arrive at the equation
\begin{align}
G^{<}_{vc}(k;E)=G^{<}_{vc,(0)}(k;E)+\mathcal{T}(k;E)\int dq\,
\tilde{V}(k,q)G^{<}_{vc}(q;E)\label{eq:scalar_BSE_bulk}
\end{align}
with the effective Coulomb potential
\begin{align}
\tilde{V}(k,q)=\frac{e^2}{4\pi^2\varepsilon\varepsilon_{0}}\frac{q}{k}\mathrm{ln}\left[\frac{k^{2}+q^{2}
+2kq+q_{0}^2}{k^{2}+q^{2}-2kq+q_{0}^2}\right].
\end{align}
The BSE in \eqref{eq:scalar_BSE_bulk} can then be rewritten as
\begin{align}
G^{<}_{vc,(0)}(k;E)=&\int dq\,\mathcal{M}(k,q)G^{<}_{vc}(q;E),
\end{align}
with
\begin{align}
\mathcal{M}(k,q)\equiv\delta(k-q)-\mathcal{T}(k;E)\tilde{V}(k,q),
\end{align}
which can be solved in discrete momentum space via inversion of matrix $\boldsymbol{\mathcal{M}}$. 
Fig.~\ref{fig:excitons_bulk}a) shows the bulk absorption coefficient 
\begin{align}
\bar{\alpha}(E_{\gamma})=\frac{\hbar}{\varepsilon_{0}c_{0}\sqrt{\varepsilon_{b}}
E_{\gamma}}\left(\frac{e}{m_{0}}\right)^2\bar{p}_{cv}^2\sum_{\mathbf{k}}G^{<}_{vc}(\mathbf{k};E_{\gamma})
\end{align} 
for a two band effective mass model of a direct semiconductor
with and without Coulomb correlations as derived via the macroscopic susceptibility from the coherent polarization
function given in \eqref{eq:macroscopic_interbandpol}. The parameters used in the simulation
are given in Tab.~\ref{tab:band_parameters} and with exception of the hole mass correspond to GaAs.
The inverse screening length is taken at $q_{0}=10^{6}m^{-1}$.

\begin{table}[!b]
\caption{\label{tab:band_parameters} Material parameters used in the simulations}
\begin{ruledtabular}
\begin{tabular}{cccccc}
 ~&$m^{*}_{c}/m_{0}$& $m^{*}_{v}/m_{0}$&$E_{g}$&$\bar{p}_{cv}^2/m_{0}$ &$\varepsilon$\\\hline
 \\GaAs&0.067& 0.1& 1.42 eV& 18 eV & 13.6\\
 Al$_{x}$Ga$_{1-x}$As&0.095& 0.1& 1.82 eV& 18 eV & 12.2\\
\end{tabular}
\end{ruledtabular}
\end{table} 

The expression for the (singular) interband self-energy may now be rewritten using the above result
\eqref{eq:BSE_bulk} for the interband Green's function,
\begin{align}
\Sigma^{\delta}_{vc}(\mathbf{k},E)=&\Sigma^{\delta,e\gamma}_{vc}(\mathbf{k},E)+\Sigma^{\delta,Cb}_{vc}(\mathbf{k},E)\nonumber\\
=&\Sigma^{\delta}_{vc,(0)}(\mathbf{k},E)+i\hbar\sum_{\mathbf{q}}V(\mathbf{k}-\mathbf{q})G^{<}_{vc}(\mathbf{q};E)\nonumber\\
=&\Sigma^{\delta}_{vc,(0)}(\mathbf{k},E)+\sum_{\mathbf{q}}V(\mathbf{k}-\mathbf{q})\mathcal{T}(\mathbf{q};E)
\Sigma^{\delta}_{vc}(\mathbf{q},E),\label{eq:bse_se_realspace}
\end{align}
In order to obtain the Coulomb enhancement factor for the effective interband coupling, the equation
is formulated for the normalized self-energy $\sigma^{\delta}_{vc}(\mathbf{k},E)\equiv
[\Sigma^{\delta}_{vc,(0)}(E)]^{-1}\Sigma^{\delta}_{vc}(\mathbf{k},E)$, neglecting
the quasi-momentum dependence of the momentum matrix elements,
\begin{align}
\sigma^{\delta}_{vc}(\mathbf{k},E)
=&1+\sum_{\mathbf{q}}V(\mathbf{k}-\mathbf{q})\mathcal{T}(\mathbf{q};E)\sigma^{\delta}_{vc}(\mathbf{q},E),
\end{align}
which is idependent of the exciting field an hence related to the macroscopic interband
susceptibility. The effective interband self-energy for monochromatic illumination with frequency
$\omega_{0}$ can then be written as follows:
\begin{align}
\tilde{\Sigma}_{cc}^{<,e\gamma}(\mathbf{k};E)=&|\Sigma^{\delta}_{cv,(0)}(\hbar\omega_{0})|^{2}
|\sigma^{\delta}_{vc}(\mathbf{k},\hbar\omega_{0})|^{2}\nonumber\\&\times
\tilde{G}_{vv}^{<}(\mathbf{k};E-\hbar\omega_{0}).
\end{align}

The spectral response displayed in Fig. 
\ref{fig:excitons_bulk}b)  is obtained from the photocurrent given by the bulk version of
\eqref{eq:mw},
\begin{align}
J^{\gamma}(\hbar\omega_{\gamma})=&\sum_{\alpha}\int\frac{dE}{2\pi}\sum_{\mathbf{k}}
\Big[\hbar^{-1} \Gamma^{\alpha}(\mathbf{k};E)\{n_{F}(E-\mu_{\alpha})\nonumber\\&\times 
A(\mathbf{k};E)+i G^{<}(\mathbf{k};E)\}\Big],\label{eq:mw_bulk}
\end{align}
where the Coulomb corrections enter via the effective interband self-energy \eqref{eq:eff_se} used in the equations for
the Green's functions $G^{\alpha}$, $\alpha=R/A,\lessgtr$.

\subsection{Thin films}
\begin{figure}[!b]
\includegraphics[width=8cm]{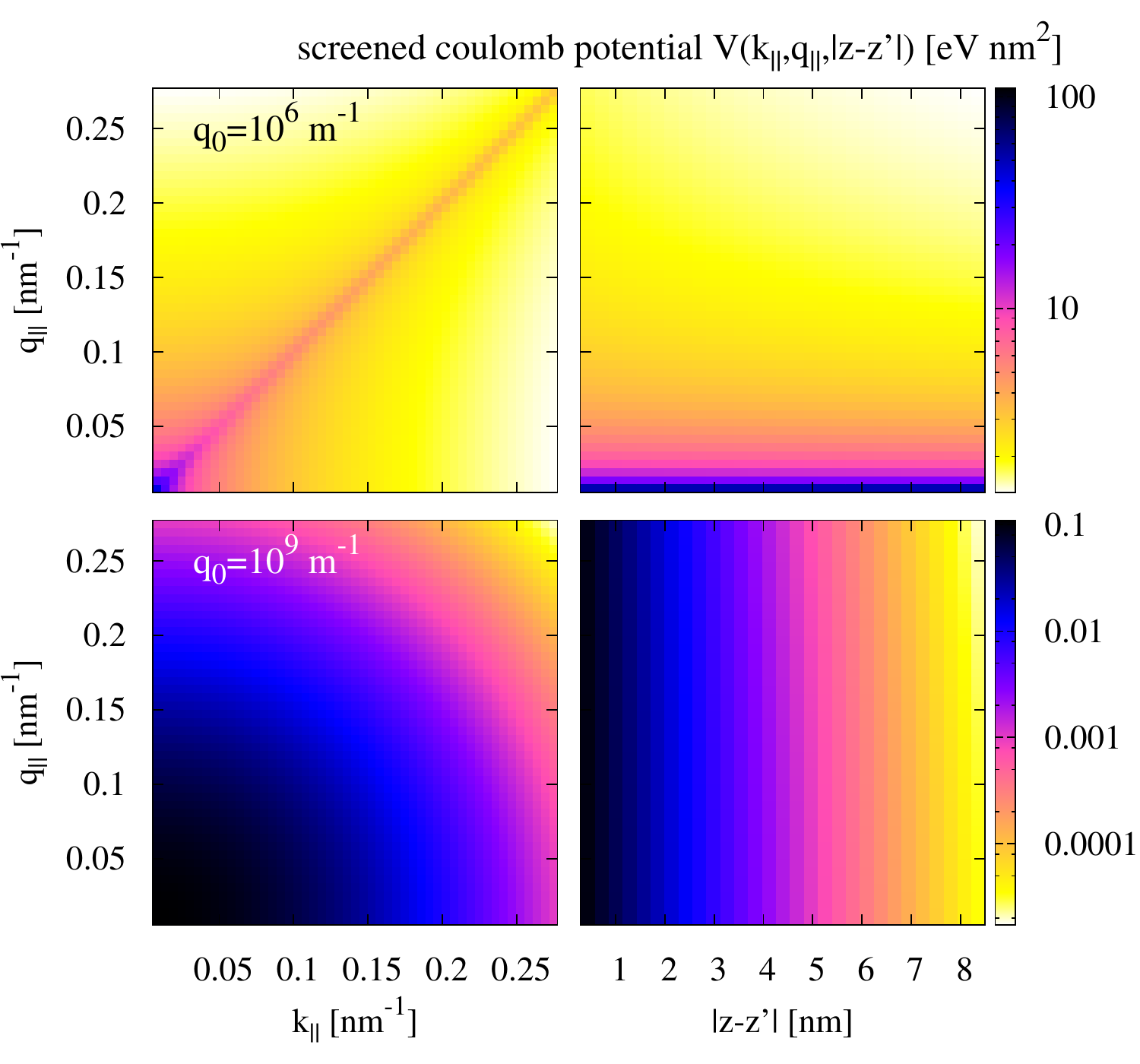}
\caption{(Color online) Quasi-momentum ($z=z'$, left column) and spatial
($q_{\parallel}=k_{\parallel}$, right column) dependence of the statically screened Coulomb
interaction for real space resolution in one dimension. In the low-screening limit ($q_{0}=10^6$
m$^{-1}$, upper row), the interaction is local in momentum space and uniform in real space. In the
strong screening regime ($q_{0}=10^9$ m$^{-1}$, lower row), the situation is reversed.
\label{fig:coulomb}}
\end{figure}

If periodicity is restricted to the transverse dimensions, Eq.
\eqref{eq:BSE_realspace} becomes 
\begin{align}
G^{<}_{vc}&(\mathbf{k}_{\parallel},z_{1},z_{1}';E)=G^{<}_{vc,(0)}(\mathbf{k}_{\parallel},z_{1},z_{1}';E)\nonumber\\&+
\int dz_{2}\int dz_{3}
\mathcal{T}(\mathbf{k}_{\parallel},z_{1},z_{2},z_{3},z_{1'};E)\nonumber\\&\times\sum_{\mathbf{q}_{\parallel}}
V(\mathbf{k}_{\parallel}-\mathbf{q}_{\parallel},z_{2},z_{3})G^{<}_{vc}(\mathbf{q}_{\parallel},z_{2},z_{3};E),\label{eq:BSE_thinfilm}
\end{align}
where $V$ is again the screened Coulomb potential. With the approximation of angular isotropy in the
transverse dimensions, the BSE equation may be written
\begin{align}
G^{<}_{vc}&(k_{\parallel},z_{1},z_{1}';E)=G^{<}_{vc,(0)}(k_{\parallel},z_{1},z_{1}';E)\nonumber\\&+
\int dz_{2}\int dz_{3}
\mathcal{T}(k_{\parallel},z_{1},z_{2},z_{3},z_{1'};E)\nonumber\\&\times\int dq_{\parallel}
\tilde{V}(k_{\parallel},q_{\parallel},z_{2},z_{3})G^{<}_{vc}(q_{\parallel},z_{2},z_{3};E),\label{eq:BSE_thinfilm_iso}
\end{align}
with the effective, statically screened Coulomb potential
\begin{align}
\tilde{V}(k_{\parallel},q_{\parallel},z_{1},z_{2})=&\frac{e^{2}}{8\pi^2\varepsilon\varepsilon_{0}}q_{\parallel}\int_{0}^{2\pi}
d\theta~
\frac{e^{-\tilde{q}(k_{\parallel},q_{\parallel},\theta)|z_{1}-z_{2}|}}{\tilde{q}(k_{\parallel},q_{\parallel},\theta)},\nonumber\\
\tilde{q}(k_{\parallel},q_{\parallel},\theta)\equiv&\sqrt{k_{\parallel}^2+q_{\parallel}^2-2k_{\parallel}q_{\parallel}\cos\theta+q_{0}^{2}}.
\end{align}
Neglecting the short-range contribution of the Bloch functions, the  Coulomb-matrix element for a localized real-space
basis set may be approximated as $\tilde{V}_{ij}(k_{\parallel},q_{\parallel})\approx
\tilde{V}(k_{\parallel},q_{\parallel},z_{i},z_{j})$. Fig.~\ref{fig:coulomb} shows the spatial and
momentum dependence of the effective, statically screened Coulomb potential for two different values of the inverse screening
length, $q_{0}=10^6$ m$^{-1}$ and $q_{0}=10^9$ m$^{-1}$, corresponding to the limiting cases of
weak and strong screening, respectively. As is to be expected, a small screening length leads to a
long-range interaction that is strongly localized in momentum space, with the opposite behavior in
the case of large screening length.

Using the localized basis representation, the BSE for the coherent interband polarization
function may be expressed as a matrix equation in analogy to the bulk case,
\begin{align}
[G^{<}_{vc}(E)]_{\alpha}=[G^{<}_{vc,0}(E)]_{\alpha}+\sum_{\beta}\tilde{M}_{\alpha\beta}(E)[G^{<}_{vc}(E)]_{\beta},\label{eq:bse_pol_tf}
\end{align}
where multi-index notation is used, with $\alpha=(i,j,k_{\parallel})$ and
$\beta=(l,m,q_{\parallel})$, and
\begin{align}
\tilde{M}_{\alpha(ijk_{\parallel}),\beta(lmq_{\parallel})}(E)=\mathcal{T}_{ilmj}(k_{\parallel};E)
\tilde{V}_{lm}(k_{\parallel},q_{\parallel})\Delta q_{\parallel}.
\end{align}
The corresponding equation for the linear susceptibility $\chi$ may be obtained from
the above equation 
\begin{align}
[\chi_{vc}(E)]_{\alpha}=[\chi_{vc,(0)}(E)]_{\alpha}+\sum_{\beta}\tilde{M}_{\alpha\beta}(E)[\chi_{vc}(E)]_{\beta},
\end{align}
with $\chi_{vc,(0)}=\mathcal{T}$. This leads to the equation
\begin{align}
\left[\mathbbm{1}-\tilde{\mathbf{M}}(E)\right]\boldsymbol{\chi}_{vc}(E)=\tilde{\boldsymbol{\chi}}_{vc,(0)}(E),
\end{align}
where
\begin{align}
[\tilde{\chi}_{vc}(E)]_{\alpha(ijk_{\parallel})}=\sum_{l}\chi_{vc,illj}(k_{\parallel};E).
\end{align}
The expression corresponding to \eqref{eq:bse_se_realspace} for the self-consistent equation
for the singular self-energy reads
\begin{align}
\Sigma_{vc}^{\delta}(\mathbf{k}_{\parallel},z_{1},z_{1}';E)&=\Sigma_{vc,(0)}^{\delta}(\mathbf{k}_{\parallel},z_{1},z_{1}';E)
\nonumber\\
&+\sum_{\mathbf{q}_{\parallel}}V(\mathbf{k}_{\parallel}-\mathbf{q}_{\parallel},z_{1}-z_{1}')\nonumber\\&\times\int
dz_{2}\int dz_{3}\mathcal{T}(\mathbf{q}_{\parallel},z_{1},z_{2},z_{3},z_{1}';E)\nonumber\\
&\times\Sigma_{vc}^{\delta}(\mathbf{q}_{\parallel},z_{2},z_{3};E).
\label{eq:sig_thinfilm}
\end{align}
In the discrete basis, the angular isotropy limit of the above equation is
\begin{align}
[\Sigma_{vc,(0)}^{\delta}]_{ij}(k_{\parallel};E)=&\sum_{lmq_{\parallel}}\Big[\delta_{lmq_{\parallel},ijk_{\parallel}}-\tilde{V}_{ij}(k_{\parallel},q_{\parallel})
\nonumber\\&\times
\mathcal{T}_{ilmj}(q_{\parallel};E)\Big][\Sigma_{vc}^{\delta}]_{lm}(q_{\parallel};E),\label{eq:bse_sig_tf}
\end{align}
which can again be solved via matrix inversion.

\begin{figure}[t] 
\includegraphics[width=8.5cm]{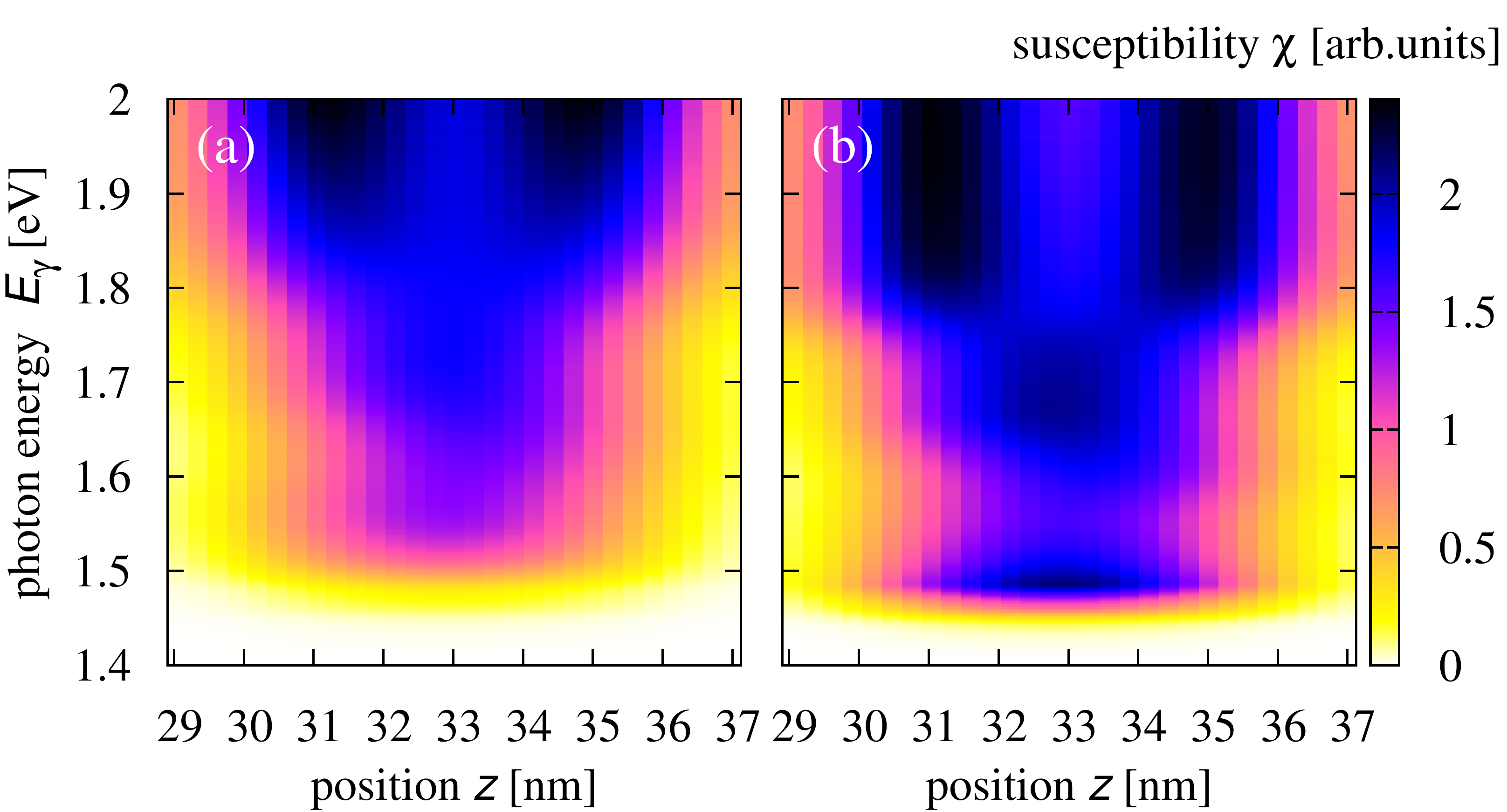}
\caption{(Color online) a) Coherent local susceptibility in the QW region for the case of
non-interacting electron-hole pairs.
b) Same for the system with excitonic corrections,
exhibiting the salient features of correlated transtions, with exciton peaks broadenend by
electron-phonon interaction and an enhancement of the quasi-continuum.
\label{fig:polfun_spectral}} 
\end{figure}

In general, the size of the matrices appearing in Eqs. \eqref{eq:bse_pol_tf} and \eqref{eq:bse_sig_tf} prohibits
the computation of the full matrix.  In a first approximation, off-diagonal elements in the spatial indices $(i,j)$ are
neglected, the long-range contributions of the electron-hole Coulomb interaction are thus
lost, which may lead to an underestimation of the exciton binding energy. The approximation is reasonable in the case of
strong screening, especially for the polarization function, where a local interaction removes the second spatial integration.
In the equation for the singular self-energy, only the diagonal elements are modified by the Coulomb interaction if the
potential is local. In the other extreme of low screening, where the Coulomb potential is spatially constant, one
still needs to account for the non-locality of $\mathcal{T}$. In the following numerical examples, the spatial
integration over $z_{3}$ in \eqref{eq:BSE_thinfilm} is considered via replacing the interaction potential with the term
$\bar{V}(k_{\parallel},q_{\parallel};z_{2})\equiv \int dz_{3}\tilde{V}(k_{\parallel},q_{\parallel};z_{2},z_{3})$. For
consistency, the same correction factor is used for the effective potential in 
\eqref{eq:sig_thinfilm}. 
\begin{figure}[!t]
\includegraphics[height=5cm]{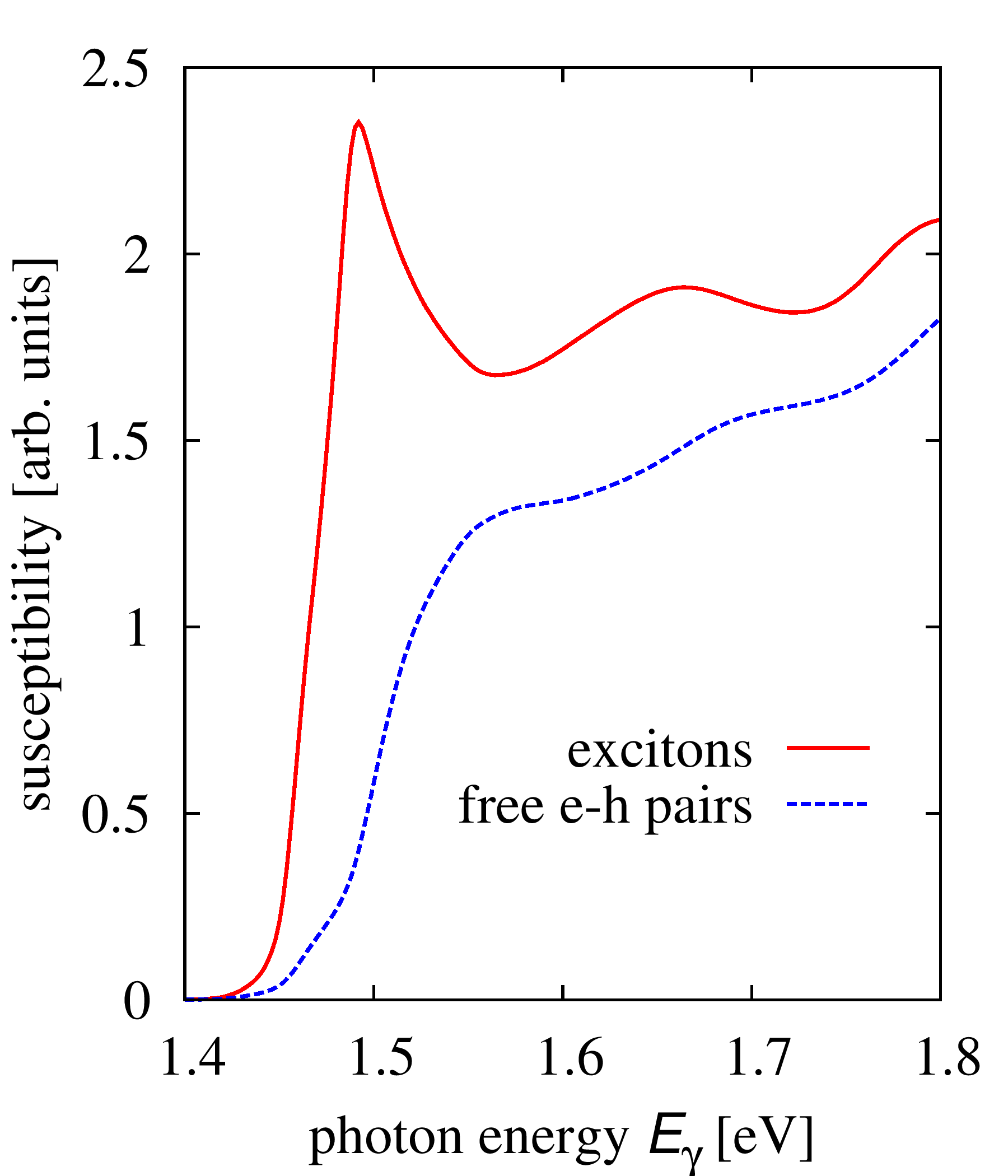}\includegraphics[height=5cm]{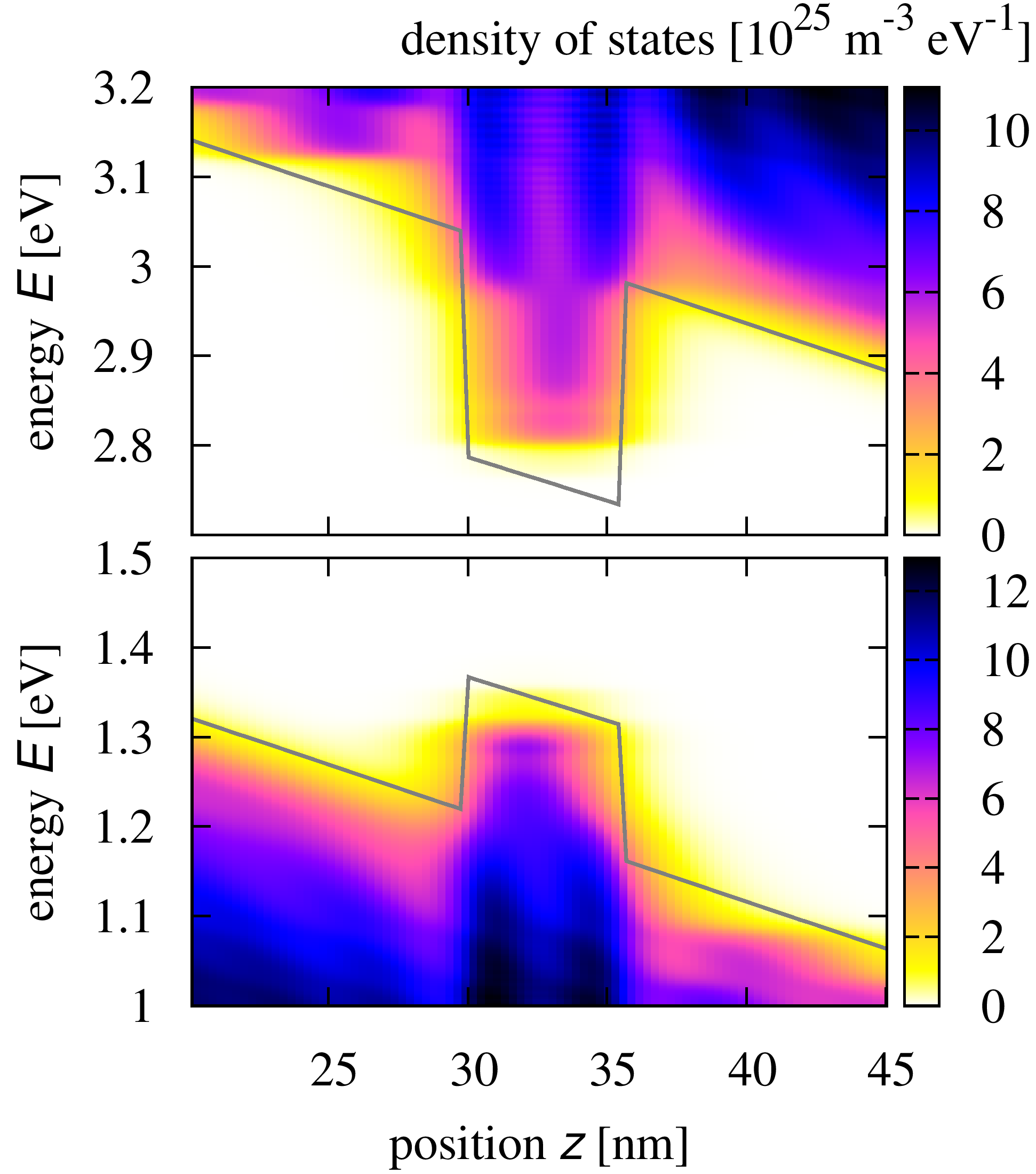}
\caption{(Color online) a) Coherent excitonic absorption and spectral response of a 5 nm GaAs
quantum well embedded in the intrinsic region of a Al$_{x}$Ga$_{1-x}$As pin diode. b) Local density
of states of the same system, revealing the situation at finite field and substantial scattering.
\label{fig:exabs_tf}} 
\end{figure}
Fig.~\ref{fig:exabs_tf}a) shows the effects of the Coulomb correlations on
absorption and photocurrent response of a 5 nm wide GaAs quantum well embedded in the
center of the intrinsic region of an Al$_{x}$Ga$_{1-x}$As ($x\sim 0.3$) $p$-$i$-$n$ diode at a
contact Fermi level splitting of 1~V and for $q_{0}=10^{6}$ m$^{-1}$. The parameters for the bulk
materials are given in Tab.~\ref{tab:band_parameters}, and the band offsets are $\Delta E_{c}=0.2$
eV and $\Delta E_{v}=0.15$. The band bending is obtained from self-consistent coupling to Poisson's
equation. Scattering is treated as in Ref.~\onlinecite{ae:prb_08}. The LDOS of the quantum well
region displayed in Fig.~\ref{fig:exabs_tf}b) reflects the effects of the sizable built-in field of
$\sim 100$ kV/cm and the strong electron-phonon interaction. The corrected susceptibility shows a distinct
exciton peak broadened by phonons. The spatially resolved susceptibility in the QW, proportional to
the local generation rate, is shown for free-electron hole pairs in Fig.~\ref{fig:polfun_spectral}a)
and for excitons in Fig.~\ref{fig:polfun_spectral}b). Again, both the appearance of the exciton
peaks below the absorption edges of the non-interacting system as well as the enhancement of the 
quasi-continuum absorption represent the salient features of the correlated transitions.

\section{Conclusions}
In this paper, a consistent inclusion of excitonic effects into the computation
of the photocurrent response of photovoltaic nanostructures was presented. While  
a full treatment of the two particle interactions is still out of range in the
context of quantum transport simulations, the excitonic enhancement of the
coupling to classical radiation fields can be considered via the corresponding
modification of the electron-photon self-energy entering the equations for the
charge carrier non-equilibrium Green's functions. However, since the
correlations near the band gap do also have a strong impact on any interband 
recombination process, it remains desirable to extend the theoretical treatment
to incoherent excitons formed by electronically or optically injected carriers.
 
\section*{Acknowledgements}
\noindent Financial support was provided by the German Federal Ministry
of Education and Research (BMBF) under Grant No. 03SF0352E.

\appendix
\section{Band decoupling procedure\label{app:decoupling}}

The standard real-time decomposition rules
\cite{langreth:76} applied to \eqref{eq:diffdys1} yield the coupled equations
for the retarded components of the intra- and interband Green's functions, 
\begin{align}
G^{-1}_{0,cc}(1,1)G^{R}_{cc}(1,1')=&\delta(1,1')
+\int\d
2\,\Big[\Sigma_{cv}^{R}(1,2)
G_{vc}^{R}(2,1')\nonumber\\&+\Sigma_{cc}^{R}(1,2)G^{R}_{cc}(2,1')\Big],\label{eq:GRcc}\\
G^{-1}_{0,vv}(1,1)G^{R}_{vc}(1,1')=&\int\d
2\,\Big[\Sigma_{vc}^{\delta}(1,2)
G_{cc}^{R}(2,1')\nonumber\\&+\Sigma_{vv}^{R}(1,2)G^{R}_{vc}(2,1')\Big].\label{eq:GRvc}
\end{align}
Introducing the new quantity 
\begin{align}
\tilde{G}_{vv}^R\equiv\left[G^{-1}_{0,vv}-\Sigma_{vv}^{R}\right]^{-1}, 
\end{align}
in \eqref{eq:GRvc}, the retarded interband Green's function can be written as
\begin{align}
G_{vc}^{R}(1,1')=\int\d
2\int \d3\,\tilde{G}_{vv}^R(1,2)\Sigma_{vc}^{\delta}(2,3)
G_{cc}^{R}(3,1').
\end{align}
Inserting the above expression in \eqref{eq:GRcc} provides a closed equation for
the retarded intraband Green's function,
\begin{align}
G_{cc}^{R}(1,1')=&\left[G^{-1}_{0,cc}(1,1')
-\Sigma_{cc}^{R}(1,1')-\Sigma_{cc}^{\delta
R}(1,1')\right]^{-1}\\
 \equiv&\left[G^{-1}_{0,cc}(1,1')-\tilde{\Sigma}_{cc}^{R}(1,
 1')\right]^{-1},\label{eq:dyson_gr}
\end{align}
where the contribution from the singular
terms to effective band-coupling intra-band self-energy $\tilde{\Sigma}$ is
\begin{align}
\Sigma_{cc}^{\delta
R}(1,1')\equiv\int \d 2\int
\d 3\,\Sigma_{cv}^{\delta}(1,2)\tilde{G}_{vv}^R(2,3)
\Sigma_{vc}^{\delta}(3,1').
\end{align}
In the same way, the lesser and greater components of the Green's functions can
be decoupled: starting from
\begin{widetext}
\begin{align}
&G^{-1}_{0,cc}(1,1)G^{<}_{cc}(1,1')=\int\d
2\,\left[\Sigma_{cv}^{\delta}(1,2)
G_{vc}^{<}(2,1')+\Sigma_{cc}^{R}(1,2)G^{<}_{cc}(2,1')
+\Sigma_{cc}^{<}(1,2)G^{A}_{cc}(2,1')\right],\label{eq:Gincc}\\
&G^{-1}_{0,vv}(1,1)G^{<}_{vc}(1,1')=\int\d2\,\left[\Sigma_{vc}^{\delta}(1,2)
G_{cc}^{<}(2,1')+\Sigma_{vv}^{R}(1,2)G^{<}_{vc}(2,1')+
\Sigma_{vv}^{<}(1,2)G^{A}_{vc}(2,1')\right].\label{eq:GRinvc}
\end{align}
\end{widetext}
the interband correlation or coherent polarization function is written as
\begin{align}
G_{vc}^{<}(1,1')=&\int\d
2\,\int\d 3\,\Big[\tilde{G}_{vv}^{R}(1,2)\Sigma_{vc}^{\delta}(2,3)
G_{cc}^{<}(3,1')\nonumber\\&+\tilde{G}_{vv}^{<}(1,2)\Sigma_{vc}^{\delta}(2,3)
G_{cc}^{A}(3,1')\Big],
\end{align}
where 
\begin{align}
\tilde{G}_{vv}^{<}(1,1')\equiv\int\d
2\int\d 3\,
\tilde{G}_{vv}^{R}(1,2)\Sigma_{vv}^{<}(2,3)
\tilde{G}_{vv}^{A}(3,1')
\end{align}
was introduced. Replacing the interband term in \eqref{eq:Gincc} then yields
the intraband correlation function 

\begin{align}
G_{cc}^{<}(1,1')=&\int\d
2\int\d
3\,
G_{cc}^{R}(1,2)\Big[\Sigma_{cc}^{<}(2,3)\nonumber\\
&+\Sigma_{cc}^{\delta<}(2,3)\Big]G_{cc}^{A}(3,1')\label{eq:keldysh_gn}\\
\equiv&\int\d 2\int\d 3\,
G_{cc}^{R}(1,2)\tilde{\Sigma}_{cc}^{<}(2,3)G_{cc}^{A}(3,1')
\end{align} 
with 
\begin{align} 
\Sigma_{cc}^{\delta<}(1,1')\equiv\int\d 2\int\d 3\,
\Sigma_{cv}^{\delta}(1,2)\tilde{G}_{vv}^{<}(2,3)
\Sigma_{vc}^{\delta}(3,1').
\end{align}
The expressions for the valence band self-energy corrections are
obtained from analogous derivations as and are identical to the above result with $c\leftrightarrow
v$.

\bibliographystyle{apsrev4-1}
\bibliography{/home/aeberurs/Biblio/bib_files/negf,/home/aeberurs/Biblio/bib_files/aeberurs,/home/aeberurs/Biblio/bib_files/scqmoptics,/home/aeberurs/Biblio/bib_files/generation,/home/aeberurs/Biblio/bib_files/pv,/home/aeberurs/Biblio/bib_files/qwsc}

\begin{thebibliography}{26}%
\makeatletter
\providecommand \@ifxundefined [1]{%
 \@ifx{#1\undefined}
}%
\providecommand \@ifnum [1]{%
 \ifnum #1\expandafter \@firstoftwo
 \else \expandafter \@secondoftwo
 \fi
}%
\providecommand \@ifx [1]{%
 \ifx #1\expandafter \@firstoftwo
 \else \expandafter \@secondoftwo
 \fi
}%
\providecommand \natexlab [1]{#1}%
\providecommand \enquote  [1]{``#1''}%
\providecommand \bibnamefont  [1]{#1}%
\providecommand \bibfnamefont [1]{#1}%
\providecommand \citenamefont [1]{#1}%
\providecommand \href@noop [0]{\@secondoftwo}%
\providecommand \href [0]{\begingroup \@sanitize@url \@href}%
\providecommand \@href[1]{\@@startlink{#1}\@@href}%
\providecommand \@@href[1]{\endgroup#1\@@endlink}%
\providecommand \@sanitize@url [0]{\catcode `\\12\catcode `\$12\catcode
  `\&12\catcode `\#12\catcode `\^12\catcode `\_12\catcode `\%12\relax}%
\providecommand \@@startlink[1]{}%
\providecommand \@@endlink[0]{}%
\providecommand \url  [0]{\begingroup\@sanitize@url \@url }%
\providecommand \@url [1]{\endgroup\@href {#1}{\urlprefix }}%
\providecommand \urlprefix  [0]{URL }%
\providecommand \Eprint [0]{\href }%
\@ifxundefined \urlstyle {%
  \providecommand \doi  [0]{\begingroup \@sanitize@url \@doi}%
  \providecommand \@doi [1]{\endgroup \@@startlink {\doibase
  #1}doi:\discretionary {}{}{}#1\@@endlink }%
}{%
  \providecommand \doi  [0]{doi:\discretionary{}{}{}\begingroup
  \urlstyle{rm}\Url }%
}%
\providecommand \doibase [0]{http://dx.doi.org/}%
\providecommand \Doi [0]{\begingroup \@sanitize@url \@Doi }%
\providecommand \@Doi  [1]{\endgroup\@@startlink{\doibase#1}\@@Doi}%
\providecommand \@@Doi [1]{#1\@@endlink}%
\providecommand \selectlanguage [0]{\@gobble}%
\providecommand \bibinfo  [0]{\@secondoftwo}%
\providecommand \bibfield  [0]{\@secondoftwo}%
\providecommand \translation [1]{[#1]}%
\providecommand \BibitemOpen [0]{}%
\providecommand \bibitemStop [0]{}%
\providecommand \bibitemNoStop [0]{.\EOS\space}%
\providecommand \EOS [0]{\spacefactor3000\relax}%
\providecommand \BibitemShut  [1]{\csname bibitem#1\endcsname}%
\bibitem [{\citenamefont {Ekins-Daukes}\ \emph {et~al.}(1999)\citenamefont
  {Ekins-Daukes}, \citenamefont {Barnham}, \citenamefont {Connolly},
  \citenamefont {Roberts}, \citenamefont {Clark}, \citenamefont {Hill},\ and\
  \citenamefont {Mazzer}}]{ned:99_2}%
  \BibitemOpen
  \bibfield  {author} {\bibinfo {author} {\bibfnamefont {N.~J.}\ \bibnamefont
  {Ekins-Daukes}}, \bibinfo {author} {\bibfnamefont {K.~W.~J.}\ \bibnamefont
  {Barnham}}, \bibinfo {author} {\bibfnamefont {J.~P.}\ \bibnamefont
  {Connolly}}, \bibinfo {author} {\bibfnamefont {J.~S.}\ \bibnamefont
  {Roberts}}, \bibinfo {author} {\bibfnamefont {J.~C.}\ \bibnamefont {Clark}},
  \bibinfo {author} {\bibfnamefont {G.}~\bibnamefont {Hill}}, \ and\ \bibinfo
  {author} {\bibfnamefont {M.}~\bibnamefont {Mazzer}},\ }\Doi
  {10.1063/1.125580} {\bibfield  {journal} {\bibinfo  {journal} {Appl. Phys.
  Lett.},\ }\textbf {\bibinfo {volume} {75}},\ \bibinfo {pages} {4195}
  (\bibinfo {year} {1999})}\BibitemShut {NoStop}%
\bibitem [{\citenamefont {Green}(2000)}]{green:00}%
  \BibitemOpen
  \bibfield  {author} {\bibinfo {author} {\bibfnamefont {M.~A.}\ \bibnamefont
  {Green}},\ }\href@noop {} {\bibfield  {journal} {\bibinfo  {journal} {J.
  Mater. Sci. Eng. B},\ }\textbf {\bibinfo {volume} {74}},\ \bibinfo {pages}
  {118 } (\bibinfo {year} {2000})}\BibitemShut {NoStop}%
\bibitem [{\citenamefont {Conibeer}\ \emph {et~al.}(2006)\citenamefont
  {Conibeer}, \citenamefont {Green}, \citenamefont {Corkish}, \citenamefont
  {Cho}, \citenamefont {Cho}, \citenamefont {Jiang}, \citenamefont
  {Fangsuwannarak}, \citenamefont {Pink}, \citenamefont {Huang}, \citenamefont
  {Puzzer}, \citenamefont {Trupke}, \citenamefont {Richards}, \citenamefont
  {Shalav},\ and\ \citenamefont {Lin}}]{conibeer:06_tsf}%
  \BibitemOpen
  \bibfield  {author} {\bibinfo {author} {\bibfnamefont {G.}~\bibnamefont
  {Conibeer}}, \bibinfo {author} {\bibfnamefont {M.}~\bibnamefont {Green}},
  \bibinfo {author} {\bibfnamefont {R.}~\bibnamefont {Corkish}}, \bibinfo
  {author} {\bibfnamefont {Y.}~\bibnamefont {Cho}}, \bibinfo {author}
  {\bibfnamefont {E.~C.}\ \bibnamefont {Cho}}, \bibinfo {author} {\bibfnamefont
  {C.~W.}\ \bibnamefont {Jiang}}, \bibinfo {author} {\bibfnamefont
  {T.}~\bibnamefont {Fangsuwannarak}}, \bibinfo {author} {\bibfnamefont
  {E.}~\bibnamefont {Pink}}, \bibinfo {author} {\bibfnamefont {Y.~D.}\
  \bibnamefont {Huang}}, \bibinfo {author} {\bibfnamefont {T.}~\bibnamefont
  {Puzzer}}, \bibinfo {author} {\bibfnamefont {T.}~\bibnamefont {Trupke}},
  \bibinfo {author} {\bibfnamefont {B.}~\bibnamefont {Richards}}, \bibinfo
  {author} {\bibfnamefont {A.}~\bibnamefont {Shalav}}, \ and\ \bibinfo {author}
  {\bibfnamefont {K.~L.}\ \bibnamefont {Lin}},\ }\href@noop {} {\bibfield
  {journal} {\bibinfo  {journal} {Thin Solid Films},\ }\textbf {\bibinfo
  {volume} {511}},\ \bibinfo {pages} {654} (\bibinfo {year}
  {2006})}\BibitemShut {NoStop}%
\bibitem [{\citenamefont {Mart\'i}\ \emph {et~al.}(2006)\citenamefont
  {Mart\'i}, \citenamefont {L\'opez}, \citenamefont {Antol\'in}, \citenamefont
  {C\'anovas}, \citenamefont {Farmer}, \citenamefont {~},\ and\ \citenamefont
  {Luque}}]{marti:06}%
  \BibitemOpen
  \bibfield  {author} {\bibinfo {author} {\bibfnamefont {A.}~\bibnamefont
  {Mart\'i}}, \bibinfo {author} {\bibfnamefont {N.}~\bibnamefont {L\'opez}},
  \bibinfo {author} {\bibfnamefont {E.}~\bibnamefont {Antol\'in}}, \bibinfo
  {author} {\bibfnamefont {S.~C.}\ \bibnamefont {C\'anovas}, \bibfnamefont
  {E.~and}}, \bibinfo {author} {\bibfnamefont {C.}~\bibnamefont {Farmer}},
  \bibinfo {author} {\bibfnamefont {L.}~\bibnamefont {~}, \bibfnamefont
  {Cuadra}}, \ and\ \bibinfo {author} {\bibfnamefont {A.}~\bibnamefont
  {Luque}},\ }\href@noop {} {\bibfield  {journal} {\bibinfo  {journal} {Thin
  Solid Films},\ }\textbf {\bibinfo {volume} {511}},\ \bibinfo {pages}
  {638} (\bibinfo {year} {2006})}\BibitemShut {NoStop}%
\bibitem [{\citenamefont {Aeberhard}\ and\ \citenamefont
  {Morf}(2008)}]{ae:prb_08}%
  \BibitemOpen
  \bibfield  {author} {\bibinfo {author} {\bibfnamefont {U.}~\bibnamefont
  {Aeberhard}}\ and\ \bibinfo {author} {\bibfnamefont {R.~H.}\ \bibnamefont
  {Morf}},\ }\href@noop {} {\bibfield  {journal} {\bibinfo  {journal} {Phys.
  Rev. B},\ }\textbf {\bibinfo {volume} {77}},\ \bibinfo {pages} {125343}
  (\bibinfo {year} {2008})}\BibitemShut {NoStop}%
\bibitem [{\citenamefont {Aeberhard}(2011){\natexlab{a}}}]{ae:nrl_11}%
  \BibitemOpen
  \bibfield  {author} {\bibinfo {author} {\bibfnamefont {U.}~\bibnamefont
  {Aeberhard}},\ }{\bibfield  {journal}
  {\bibinfo  {journal} {Nanoscale Res. Lett.},\ }\textbf {\bibinfo {volume}
  {6}},\ \bibinfo {pages} {242} (\bibinfo {year}
  {2011}{\natexlab{a}})}\BibitemShut {NoStop}%
\bibitem [{\citenamefont {Aeberhard}(2011){\natexlab{b}}}]{ae:jcel_review}%
  \BibitemOpen
  \bibfield  {author} {\bibinfo {author} {\bibfnamefont {U.}~\bibnamefont
  {Aeberhard}},\ }\href {http://dx.doi.org/10.1007/s10825-011-0375-6}
  {\bibfield  {journal} {\bibinfo  {journal} {J. Comput.
  Electron.},\ }\textbf {\bibinfo {volume} {10}},\ \bibinfo {pages} {394}
  (\bibinfo {year} {2011}{\natexlab{b}})},\BibitemShut {NoStop}%
\bibitem [{\citenamefont {Aeberhard}(2012)}]{ae:oqe_12}%
  \BibitemOpen
  \bibfield  {author} {\bibinfo {author} {\bibfnamefont {U.}~\bibnamefont
  {Aeberhard}},\ }\href {http://dx.doi.org/10.1007/s11082-011-9529-9}
  {\bibfield  {journal} {\bibinfo  {journal} {Opt. Quantum. Electron.},\
  }\textbf {\bibinfo {volume} {44}},\ \bibinfo {pages} {133} (\bibinfo {year}
  {2012})},\BibitemShut {NoStop}%
\bibitem [{\citenamefont {Haug}\ and\ \citenamefont
  {Schmitt-Rink}(1984)}]{haug:84}%
  \BibitemOpen
  \bibfield  {author} {\bibinfo {author} {\bibfnamefont {H.}~\bibnamefont
  {Haug}}\ and\ \bibinfo {author} {\bibfnamefont {S.}~\bibnamefont
  {Schmitt-Rink}},\ }\Doi {DOI: 10.1016/0079-6727(84)90026-0} {\bibfield
  {journal} {\bibinfo  {journal} {Prog. Quantum Electron.},\ }\textbf
  {\bibinfo {volume} {9}},\ \bibinfo {pages} {3 } (\bibinfo {year} {1984})},\BibitemShut {NoStop}%
\bibitem [{\citenamefont {Haug}\ and\ \citenamefont
  {Henneberger}(1988)}]{haug:88}%
  \BibitemOpen
  \bibfield  {author} {\bibinfo {author} {\bibfnamefont {H.}~\bibnamefont
  {Haug}}\ and\ \bibinfo {author} {\bibfnamefont {K.}~\bibnamefont
  {Henneberger}},\ }\href@noop {} {\bibfield  {journal} {\bibinfo  {journal}
  {Phys. Rev. B},\ }\textbf {\bibinfo {volume} {38}},\ \bibinfo {pages} {9759}
  (\bibinfo {year} {1988})}\BibitemShut {NoStop}%
\bibitem [{\citenamefont {Haug}(1992)}]{haug:92}%
  \BibitemOpen
  \bibfield  {author} {\bibinfo {author} {\bibfnamefont {H.}~\bibnamefont
  {Haug}},\ }\href@noop {} {\bibfield  {journal} {\bibinfo  {journal} {Phys.
  Status Solidi B},\ }\textbf {\bibinfo {volume} {173}},\ \bibinfo {pages}
  {139} (\bibinfo {year} {1992})}\BibitemShut {NoStop}%
\bibitem [{\citenamefont {Haug}\ and\ \citenamefont {Jauho}(1996)}]{haug:96}%
  \BibitemOpen
  \bibfield  {author} {\bibinfo {author} {\bibfnamefont {H.}~\bibnamefont
  {Haug}}\ and\ \bibinfo {author} {\bibfnamefont {A.~P.}\ \bibnamefont
  {Jauho}},\ }\href@noop {} {\emph {\bibinfo {title} {Quantum kinetics in
  transport and optics of semiconductors}}}\ (\bibinfo  {publisher} {Springer,
  Berlin},\ \bibinfo {year} {1996})\BibitemShut {NoStop}%
\bibitem [{\citenamefont {Haug}\ and\ \citenamefont {Koch}(2004)}]{haug:04}%
  \BibitemOpen
  \bibfield  {author} {\bibinfo {author} {\bibfnamefont {H.}~\bibnamefont
  {Haug}}\ and\ \bibinfo {author} {\bibfnamefont {S.~W.}\ \bibnamefont
  {Koch}},\ }\href@noop {} {\emph {\bibinfo {title} {Quantum Theory of the
  Optical and Electronic Properties of Semiconductors}}}\ (\bibinfo
  {publisher} {World Scientific},\ \bibinfo {year} {2004})\BibitemShut
  {NoStop}%
\bibitem [{\citenamefont {Henneberger}\ and\ \citenamefont
  {May}(1986)}]{henneberger:86}%
  \BibitemOpen
  \bibfield  {author} {\bibinfo {author} {\bibfnamefont {K.}~\bibnamefont
  {Henneberger}}\ and\ \bibinfo {author} {\bibfnamefont {V.}~\bibnamefont
  {May}},\ }\href@noop {} {\bibfield  {journal} {\bibinfo  {journal} {Physica
  A},\ }\textbf {\bibinfo {volume} {138}},\ \bibinfo {pages} {537} (\bibinfo
  {year} {1986})}\BibitemShut {NoStop}%
\bibitem [{\citenamefont {Henneberger}(1988){\natexlab{a}}}]{henneberger:88}%
  \BibitemOpen
  \bibfield  {author} {\bibinfo {author} {\bibfnamefont {K.}~\bibnamefont
  {Henneberger}},\ }\href@noop {} {\bibfield  {journal} {\bibinfo  {journal}
  {Physica A},\ }\textbf {\bibinfo {volume} {150}},\ \bibinfo {pages} {419}
  (\bibinfo {year} {1988}{\natexlab{a}})}\BibitemShut {NoStop}%
\bibitem [{\citenamefont {Henneberger}(1988){\natexlab{b}}}]{henneberger:88_2}%
  \BibitemOpen
  \bibfield  {author} {\bibinfo {author} {\bibfnamefont {K.}~\bibnamefont
  {Henneberger}},\ }\href@noop {} {\bibfield  {journal} {\bibinfo  {journal}
  {Physica A},\ }\textbf {\bibinfo {volume} {150}},\ \bibinfo {pages} {439}
  (\bibinfo {year} {1988}{\natexlab{b}})}\BibitemShut {NoStop}%
\bibitem [{\citenamefont {Henneberger}\ and\ \citenamefont
  {Haug}(1988)}]{henneberger:88_3}%
  \BibitemOpen
  \bibfield  {author} {\bibinfo {author} {\bibfnamefont {K.}~\bibnamefont
  {Henneberger}}\ and\ \bibinfo {author} {\bibfnamefont {H.}~\bibnamefont
  {Haug}},\ }\href@noop {} {\bibfield  {journal} {\bibinfo  {journal} {Phys.
  Rev. B},\ }\textbf {\bibinfo {volume} {38}},\ \bibinfo {pages} {9759}
  (\bibinfo {year} {1988})}\BibitemShut {NoStop}%
\bibitem [{\citenamefont {Henneberger}\ and\ \citenamefont
  {Koch}(1996)}]{henneberger:96}%
  \BibitemOpen
  \bibfield  {author} {\bibinfo {author} {\bibfnamefont {K.}~\bibnamefont
  {Henneberger}}\ and\ \bibinfo {author} {\bibfnamefont {S.~W.}\ \bibnamefont
  {Koch}},\ } {\bibfield  {journal} {\bibinfo
   {journal} {Phys. Rev. Lett.},\ }\textbf {\bibinfo {volume} {76}},\ \bibinfo
  {pages} {1820} (\bibinfo {year} {1996})}\BibitemShut {NoStop}%
\bibitem [{\citenamefont {Jahnke}\ and\ \citenamefont
  {Koch}(1995)}]{jahnke:95}%
  \BibitemOpen
  \bibfield  {author} {\bibinfo {author} {\bibfnamefont {F.}~\bibnamefont
  {Jahnke}}\ and\ \bibinfo {author} {\bibfnamefont {S.~W.}\ \bibnamefont
  {Koch}},\ } {\bibfield  {journal} {\bibinfo
  {journal} {Phys. Rev. A},\ }\textbf {\bibinfo {volume} {52}},\ \bibinfo
  {pages} {1712} (\bibinfo {year} {1995})}\BibitemShut {NoStop}%
\bibitem [{Note1()}]{Note1}%
  \BibitemOpen
  \bibinfo {note} {The description of spontaneous emission would require the
  additional coupling to an incoherent internal photon field.}\BibitemShut
  {NoStop}%
\bibitem [{\citenamefont {Kadanoff}\ and\ \citenamefont
  {Baym}(1962)}]{kadanoff:62}%
  \BibitemOpen
  \bibfield  {author} {\bibinfo {author} {\bibfnamefont {L.~P.}\ \bibnamefont
  {Kadanoff}}\ and\ \bibinfo {author} {\bibfnamefont {G.}~\bibnamefont
  {Baym}},\ }\href@noop {} {\emph {\bibinfo {title} {Quantum Statistical
  Mechanics}}}\ (\bibinfo  {publisher} {Benjamin, Reading, Mass.},\ \bibinfo
  {year} {1962})\BibitemShut {NoStop}%
\bibitem [{\citenamefont {Keldysh}(1965)}]{keldysh:65}%
  \BibitemOpen
  \bibfield  {author} {\bibinfo {author} {\bibfnamefont {L.}~\bibnamefont
  {Keldysh}},\ }\href@noop {} {\bibfield  {journal} {\bibinfo  {journal} {Sov.
  Phys. JETP},\ }\textbf {\bibinfo {volume} {20}},\ \bibinfo {pages} {1018}
  (\bibinfo {year} {1965})}\BibitemShut {NoStop}%
\bibitem [{\citenamefont {Langreth}(1976)}]{langreth:76}%
  \BibitemOpen
  \bibfield  {author} {\bibinfo {author} {\bibfnamefont {D.}~\bibnamefont
  {Langreth}},\ }\href@noop {} {\bibfield  {journal} {\bibinfo  {journal} {in
  \emph{Linear and Non-linear Electron Transport in solids}},\ }\textbf
  {\bibinfo {volume} {17}},\ \bibinfo {pages} {3} (\bibinfo {year}
  {1976})}\BibitemShut {NoStop}%
\bibitem [{\citenamefont {Aeberhard}(2011){\natexlab{c}}}]{ae:prb_11}%
  \BibitemOpen
  \bibfield  {author} {\bibinfo {author} {\bibfnamefont {U.}~\bibnamefont
  {Aeberhard}},\ }\Doi {10.1103/PhysRevB.84.035454} {\bibfield  {journal}
  {\bibinfo  {journal} {Phys. Rev. B},\ }\textbf {\bibinfo {volume} {84}},\
  \bibinfo {pages} {035454} (\bibinfo {year} {2011}{\natexlab{c}})}\BibitemShut
  {NoStop}%
\bibitem [{\citenamefont {Sch\"afer}\ and\ \citenamefont
  {Wegener}(2002)}]{schaefer:02}%
  \BibitemOpen
  \bibfield  {author} {\bibinfo {author} {\bibfnamefont {W.}~\bibnamefont
  {Sch\"afer}}\ and\ \bibinfo {author} {\bibfnamefont {M.}~\bibnamefont
  {Wegener}},\ }\href@noop {} {\emph {\bibinfo {title} {Semiconductor Optics
  and Transport Phenomena}}}\ (\bibinfo  {publisher} {Springer, Berlin},\
  \bibinfo {year} {2002})\BibitemShut {NoStop}%
\bibitem [{\citenamefont {Meir}\ and\ \citenamefont
  {Wingreen}(1992)}]{meir:92}%
  \BibitemOpen
  \bibfield  {author} {\bibinfo {author} {\bibfnamefont {Y.}~\bibnamefont
  {Meir}}\ and\ \bibinfo {author} {\bibfnamefont {N.}~\bibnamefont
  {Wingreen}},\ }\href@noop {} {\bibfield  {journal} {\bibinfo  {journal}
  {Phys. Rev. Lett.},\ }\textbf {\bibinfo {volume} {68}},\ \bibinfo {pages}
  {2512} (\bibinfo {year} {1992})}\BibitemShut {NoStop}%
\end{thebibliography}%

\end{document}